# Chronic iEEG recordings and interictal spike rate reveal multiscale temporal modulations in seizure states

Gabrielle M. Schroeder[1], Philippa J. Karoly[2,3], Matias Maturana[2,3,4],
Mariella Panagiotopoulou[1], Peter N. Taylor[1,5,6],
Mark J. Cook[2], Yujiang Wang[1,5,6*]

1. CNNP Lab (www.cnnp-lab.com), Interdisciplinary Computing and Complex BioSystems Group, School of Computing, Newcastle University, Newcastle upon Tyne, United Kingdom

2. Graeme Clark Institute and St Vincent's Hospital, University of Melbourne, Melbourne, Victoria, Australia

3. Department of Biomedical Engineering, University of Melbourne, Melbourne, Victoria, Australia

4. Seer Medical Pty Ltd, Melbourne, VIC, Australia

5. Faculty of Medical Sciences, Newcastle University, Newcastle upon Tyne, United Kingdom

6. UCL Queen Square Institute of Neurology, Queen Square, London, United Kingdom

* Yujiang.Wang@newcastle.ac.uk

## Abstract

Background and Objectives: Many biological processes are modulated by rhythms on circadian and multidien timescales. In focal epilepsy, various seizure features, such as spread and duration, can change from one seizure to the next within the same patient. However, the specific timescales of this variability, as well as the specific seizure characteristics that change over time, are unclear.

Methods: Here, in a cross-sectional observational study, we analysed within-patient seizure variability in 10 patients with chronic intracranial EEG recordings (185-767 days of recording time, 57-452 analysed seizures/patient). We characterised the seizure evolutions as sequences of a finite number of patient-specific functional seizure network states (SNSs). We then compared SNS occurrence and duration to (1) time since implantation and (2) patient-specific circadian and multidien cycles in interictal spike rate.

Results: In most patients, the occurrence or duration of at least one SNS was associated with the time since implantation. Some patients had one or more SNSs that were associated with phases of circadian and/or multidien spike rate cycles. A given SNS's occurrence and duration were usually not associated with the same timescale.

Discussion: Our results suggest that different time-varying factors modulate within-patient seizure evolutions over multiple timescales, with separate processes modulating a SNS's occurrence and duration. These findings imply that the development of time-adaptive treatments in epilepsy must account for several separate properties of epileptic seizures, and similar principles likely apply to other neurological conditions.

# Introduction

Focal epilepsy is characterised by recurrent, unprovoked seizures. Importantly, these seizures are not homogeneous events, even in the same patient. Within individual patients, seizure features such as clinical symptoms[1], onset locations and patterns[2–6], duration[7,8], and network evolutions[9] can change over time and potentially influence treatment responses[7,10,11]. As such, a better understanding of the patterns and sources of within-patient seizure variability is needed.

One open question is whether and how such seizure features change over short (e.g., 24h) and long (e.g., weekly, monthly, and yearly) periods of time. There is some evidence that seizures are modulated over such timescales. Certain clinical seizure types and symptoms, such as focal to bilateral tonic-clonic seizures, can preferentially occur during specific parts of sleep/wake or day/night cycles[12–16]. Electrographic seizure onset patterns can shift across the days of epilepsy monitoring unit[3] and months of chronic intracranial EEG (iEEG)[17] recordings, suggesting that seizure features can also change over slower timescales. To quantify within-patient variability in seizure dynamics, we recently compared seizure functional network evolutions[9], which capture the time-varying relationships (e.g., correlation or coherence) between the activity of different brain regions from seizure start to end. This description of seizure activity builds on the concept of epilepsy as a network disorder[18,19], and seizure network features have been linked to seizure symptoms[20–22], seizure termination[23,24], and the seizure onset zone[20,25,26]. Our preliminary analysis in epilepsy monitoring unit patients found that seizure network evolutions do not change randomly over time[9]. Instead, in most patients, the changes in seizure network evolutions could be explained by a combination of circadian and/or slower time-varying factors. However, these temporal associations and the specific timescales and seizure network changes need to be characterised in longer recordings with larger numbers of seizures.

In recent years, chronic iEEG recordings over months to years have provided unprecedented insights into epileptic brain dynamics over multiple timescales[27–30]. First, these recordings have revealed fluctuations in interictal dynamics, including in the rates and spatial patterns of bursts[17], spikes[31], high frequency activity[31], and other signal features[32]. This variability is especially high in the first months after electrode implantation, possibly due to the brain's response to acute trauma[17,31,32]. However, more persistent variability in such features has also been observed[17,31], suggesting that other mechanisms also drive the observed interictal shifts. In addition, multiple studies have found prevalent patient-specific circadian, multidien (multi-day), and/or circannual cycles in interictal features and seizure occurrence[31,33–37]. Since the exact periods of these cycles often vary over time, they are best tracked using fluctuations in continuous biomarkers such as interictal spike rate[34,35,37]. An intriguing possibility is that seizure characteristics could also change over such cycles. However, the relationship between seizure features and spike rate cycles has not been explored.

We addressed these questions by analysing changes in seizure networks in chronic iEEG recordings from the NeuroVista dataset[28]. We follow a popular approach of describing seizures as a sequence of a small number of patient-specific functional network states[20,25], each of which described a recurring relationship between the recorded brain areas. Compared to our previous descriptions of seizure network evolutions[8,9], this states-based approach allowed us to easily identify and compare which network patterns occurred in each seizure. In each patient, we then analysed changes in seizure network states over multiple timescales. We first identified gradual changes in seizure network states across the course of each recording. We then determined if seizure network states also fluctuated over patient-specific circadian and multidien cycles that were revealed by interictal spike rate. To account for possible independent variability in seizure evolutions and seizure duration[8], we separately examined variability in seizure network state occurrence and seizure network state duration. We show that in most patients, both of these features were associated with multiple timescales, providing new insight into the patterns and possible mechanisms of within-patient seizure variability.

## Methods

In the following we summarise our methods, while Supplemental Methods provide detailed descriptions of the analyses.

### Patients and seizure data

We analysed seizure data from 10 NeuroVista patients that underwent chronic iEEG recordings[28]. The seizure recordings are part of a dataset of 12 patients that was previously made available by[38]. From the original cohort, patients NeuroVista 2 and NeuroVista 4 were excluded from our analysis due to low numbers of recorded seizures (32 and 22 seizures, respectively). All other patients had at least 57 analysable seizures. The patients and collection of their chronic iEEG data is described in detail in[28], and patient details are provided in Supplementary Table S1.1. Anonymised data was analysed under the approval of the Newcastle University Ethics Committee (reference number 6887/2018). The original data acquisition is detailed in ref[28], and "the human research ethics committees of the participating institutes approved the study and subsequent amendments. All patients gave written informed consent before participation"[28].

Seizures were annotated by clinical staff after identification using patient diaries, audio recordings, and a seizure detection algorithm[7]. Seizures with clinical manifestations and corresponding iEEG changes ("type 1" seizures) and seizures with iEEG changes comparable to type 1 seizures, but without confirmed clinical manifestations ("type 2" seizures) were included in the analysis[7,38]. We excluded seizures with noisy segments (identified visually) and duration less than 10s.

## Computing progressions of seizure network states (SNSs)

After re-referencing the iEEG data to a common average reference, we obtained the sliding window (10s window, 9s overlap) functional connectivity (band-averaged coherence) in six frequency bands: delta 1-4 Hz, theta 4-8 Hz, alpha 8-13 Hz, beta 13-30 Hz, gamma 30-80 Hz, and high gamma 80-150 Hz. To characterise each patient's seizure evolutions, we soft-clustered the windows in time using non-negative matrix factorisation (NMF)[39], allowing us to assign each time window a "seizure network state" (SNS). The number of SNSs in each patient was determined using stability NMF[40]. This approach, which we have previously applied to seizure iEEG data[9], takes advantage of the nondeterministic nature of NMF to find which number of SNSs always produces the same set of SNSs. We then described seizure network evolution as a sequence, or progression, of SNSs. Note that NMF does not produce orthogonal SNSs, allowing some overlap in connectivity patterns between SNSs (see Supplementary Section S2).

## Defining SNS features

In each patient, we investigated two types of variability in SNSs: SNS occurrence and SNS duration.

SNS occurrence was a binary seizure feature defined as whether a given SNS occurred in the seizure. Most SNSs did not occur in all of a patient's seizures; thus, these SNS had variable occurrence. We excluded SNSs that occurred in all seizures from the SNS occurrence analysis.

SNS duration was a continuous measure that quantified the number of time windows a seizure spent in a given SNS. For each SNS, we analysed SNS duration in all seizures containing that SNS; in other words, zero durations were excluded from the analysis to avoid effects driven by SNS occurrence. We analysed this feature for all SNSs.

## Defining temporal features

We compared SNS features of each patient's seizures to two types of temporal features: time since implantation and spike rate cycle phase.

A seizure's time since implantation was defined as the number of days after the recording's start that the seizure occurred. This measure captured changes in SNSs over the course of the entire iEEG recording.

To analyse SNSs over shorter timescales, we first extracted fluctuations in interictal epileptiform spike rate. The continuously-recorded spike rate of each patient was detected and validated in a previous study[41]. We then extracted the prominent circadian and multidien (multi-day) fluctuations in spike rate using Empirical Mode Decomposition (EMD)[42–44] (See Supplemental Methods for details). EMD is a data-driven technique that decomposes a time series into a series of oscillatory components that can have variable amplitude and frequency over time, thus accommodating non-stationarity in the spike rate cycles. We limited our analysis to the most prominent components that had short periods relative to the length of the patient's recording. We refer to each component as a "spike rate cycle" throughout the rest of the paper. For each patient, we then extracted the phases at which each seizure occurred, thus assigning each seizure a phase for each spike rate cycle. Seizures were excluded from this analysis if they occurred during a time period with insufficient spike rate cycle data (see Supplemental Methods).

## Statistical analysis

For each patient, we compared their SNS features (SNS occurrence and duration) to each temporal feature (time since implantation and the patient's spike rate cycle phases).

For each SNS with variable occurrence, a Wilcoxon rank sum test was used to compare the time since implantation of seizures with and without the SNS. To quantify the temporal separation of seizures with and without the SNS, the area under the curve (AUC) of the receiver operating characteristic curve for distinguishing SNS occurrence using seizure times was also computed. Note that AUCs are mathematically equivalent to the Wilcoxon rank sum test statistic and therefore have the same statistical significance.

To compare SNS duration to seizure time since implantation, we computed the Spearman correlation between non-zero SNS durations and the seizure times. We used the corresponding test statistic to determine the statistical significance of the association.

To compare SNS occurrence to spike rate cycles, for each spike rate cycle we determined the phase preference of seizures with the SNS by computing the mean resultant vector from the seizure phases:

$$Re^{-i\psi} = \frac{1}{S}\sum_{s=1}^{S} e^{-i\phi_s}$$

Here, $S$ is the number of seizures containing the SNS, $s$ is a seizure with the SNS, $\phi_s$ is the spike rate cycle phase of seizure $s$, $R$ is the modulus of the mean resultant vector, and $\psi$ is the angle of the mean resultant vector. As in previous work[34,35], we refer to $R$ as the PLV of seizures containing the SNS. The PLV varies from 0 to 1 and is higher when the cycle's phases are similar across all seizures with the SNS. To control for any seizure phase preference, we used permutation tests to determine the significance of the observed PLV for each SNS and spike rate cycle. For a SNS that occurred $S$ times, we randomly selected $S$ seizures from all analysed seizures and recomputed the PLV. Repeating this process for 10,000 different permutations yielded a null distribution of PLVs for the scenario that the SNS had no additional phase preference within that spike rate cycle. The $p$-value of the SNS's phase preference was defined as the percentage of times a permuted PLV was greater than or equal to the observed PLV.

Finally, we compared SNS duration to each spike rate cycle. For each cycle and SNS, we computed the rank linear-circular correlation $D$ between the SNS's duration and the cycle phases of the corresponding seizures[45]. $D$ varies from 0 to 1, with higher values indicating a stronger association between the SNS's duration and the spike rate cycle phases. We determined the significance of these associations by randomly shuffling the seizures' SNS duration 10,000 times and computing a null distribution of correlations. The $p$-value of the observed correlation was the percentage of times a permuted correlation was greater than or equal to the observed correlation.

Benjamini-Hochberg false discovery rate (FDR) correction for multiple comparisons, with $\alpha = 0.05$, was performed for all tests across all patients that compared seizure features (SNS occurrence and SNS duration) to temporal features (seizure time in the recording and spike rate cycles), and only significant results are reported. Uncorrected $p$-values are reported in the text for reference.

### Code and data availability

All analysis was performed in MATLAB version R2018b. EMD of interictal spike rate was performed using the MATLAB package CEEMDAN (https://github.com/macolominas/CEEMDAN)[43,44]. The remaining analysis was performed using custom MATLAB scripts. The NeuroVista seizure iEEG data used in this study is available from www.epilepsyecosystem.org. The processed data (NMF $W$ and $H$ matrices) of all patients is available on Zenodo (DOI 10.5281/zenodo.5503590).

## Results

### Seizure network evolutions vary from seizure to seizure within individual patients

In each patient, we first characterized changes in seizure network evolutions over the course of each patient's continuous iEEG recording. Specifically, we described seizure network evolutions as a sequence, or progression, of seizure network states (SNSs). Fig. 1A shows the SNS progressions of two example seizures in patient NeuroVista 1. Both seizures began with the same three SNSs (B, E, and C), but had different final SNS (D in seizure 59 vs. F in seizure 65). Visually, the amplitude and frequency of the seizure activity also differ once their SNS progressions diverge. Supplementary section S2 provides the complete SNS visualisations of NeuroVista 1 to visualize the information captured by each SNS.

Fig. 1B shows all of NeuroVista 1's SNS progressions. It is already visually apparent that SNS progression differed from seizure to seizure. In particular, there was variability in both SNS occurrence (Fig. 1C) and SNS duration (Fig. 1D). For example, SNS D only occurred in 6.4 % of seizures, and, in those seizures, its duration could range from 8 to 52 windows.

Across patients, we also observed variability in SNS progressions (Fig. 1E,F, and see Zenodo Data File 10.5281/zenodo.5910238 for data for all patients). Importantly, SNS are not comparable between patients in our study due to the patient-specific intracranial implantation schemes, which record from different brain areas in each patient. Since both SNS occurrence and duration varied across all patients, we investigated both features in the following.

## Seizure network states vary over the duration of chronic iEEG recordings

We first asked if within-patient SNS occurrence and duration varied over the timescale of each patient's entire chronic iEEG recording. Specifically, we explored whether seizures that occurred early in the recording had different features from those seizures that occurred later.

We first investigated relationships between the amount of time elapsed since the iEEG implantation and SNS occurrence. Fig. 2A-B shows the relationship of an example SNS with recording time. In this patient, NeuroVista 13, only some seizures contained SNS C (Fig. 2A), and these seizures tended to occur towards the end of the recording period. The temporal separability of seizures with and without SNS C can be characterised by the AUC, with an AUC below 0.5 indicating that the SNS preferentially occurred in earlier seizures and an AUC above 0.5 revealing that the SNS tended to occur in later seizures. Here, SNS C has an AUC of 0.71, which was significant after FDR correction for multiple comparisons (Wilcoxon rank-sum test, $p = 2.1 \times 10^{-12}$).

Across our cohort, eight out of the ten patients had at least one SNS where the SNS's occurrence was significantly associated with the time since implantation (Fig. 2C). These temporal associations were not driven by transient SNSs that only occurred during the initial part of the recording (Supplementary S4).

NeuroVista 15's SNS D demonstrates how SNS duration can also vary over the length of the recording (Fig. 3A-B). Here, SNS D's duration was significantly higher in earlier seizures, as demonstrated by a significant Spearman's correlation $\rho$ of $-0.59$ after FDR correction for multiple comparisons ($p = 2.4 \times 10^{-6}$).

Almost all patients had at least one SNS where the SNS's duration was either significantly positively (eight SNSs) or negatively (eight SNSs) correlated with time since implantation (Fig. 3C-D). These findings demonstrate that, across a patient's chronic iEEG recording, it is also possible for a SNS's duration to increase or decrease.

## Seizure network states fluctuate over circadian and multidien cycles

Next, we hypothesised that SNSs, like seizure occurrence[34–37], may vary over circadian and multidien cycles. Importantly, these cycles can be nonstationary, with the cycle period varying over time; thus, they must be extracted using a continuous biomarker such as interictal spike rate[34–37].

As in previous work[34–37,41], we observed high levels of variability in interictal spike rate across each patient's chronic iEEG recording (see Fig. 4A for interictal spike rate of an example patient, NeuroVista 1). We obtained each patient's interictal spike rate from a previous study[41] and used a data-driven approach, EMD[42–44], to extract prominent spike rate cycles over different timescales (see Zenodo Data File 10.5281/zenodo.5910238 for spike rate decompositions of all patients). EMD can extract cycles with variable periods, which is a common characteristic of multidien spike rate cycles[34–37]. To approximate each spike rate cycle's period, we report each cycle's average period across the iEEG recording. Fig. 4B shows the extracted spike rate cycles of NeuroVista 1. We observed such multiscale fluctuations in interictal spike rate in most patients (Fig. 4C and D, see Supplementary section S3 for selection of spike rate timescales). All patients had prominent circadian cycles in spike rate (average period of 0.84-1.02 days), and eight of the ten patients also had at least one multidien cycle, with average periods ranging from 3.60 to 54.77 days. Together, these cycles characterised the prominent patient-specific, non-stationary changes in spike rate in each patient.

For each patient, we first asked whether a given SNS preferentially occurred during certain phases of each spike rate cycle. Similar to previous work[33–35], we defined phase preference as the PLV of a SNS for a spike rate cycle. A PLV of 0 would indicate that the SNS had no phase preference, while a PLV of 1 would indicate that the SNS only occurred at one specific phase of the cycle. Importantly, seizures themselves usually have phase preferences for circadian and multidien spike rate cycles[34–37]. Therefore, to control for seizure timing phase preferences, we used permutation tests to determine the significance of each SNS's PLV. In other words, we determined if the SNS's phase preference was significantly higher than the phase preference of the patient's seizures.

Fig. 5A-D shows an example SNS, SNS F, that preferentially occurred during certain phases of NeuroVista 1's multidien spike rate cycle. In this example, SNS F was most likely to occur during a specific part of the rising phase of the multidien cycle, with the proportion of seizures with this SNS tapering towards the cycle peak. Further, almost all seizures that occurred during the falling phase and the cycle trough lacked this SNS. As such, SNS F's PLV was significantly stronger than the overall seizure PLV (SNS PLV = 0.84, +0.21 relative to PLV of all seizures, $p = 0.0014$) after FDR correction for multiple comparisons.

Across our cohort, four patients (NeuroVista 1, 7, 10, and 13) had at least one SNS occurrence that was significantly associated spike rate cycle (Fig. 5E-F). The same SNS could be associated with multiple different spike rate cycles (see Supplementary section S5). Four of these patients had a SNS that had a phase preference in their circadian cycle, while two patients had one or more SNS associated with at least one multidien cycle (Fig. 5G). The effect-increase in PLV varied from 0.07 to 0.22 (median: 0.12, Fig. 5H). We interpret these associations as evidence that, in some patients, certain spike rate cycles reveal a modulation in SNS occurrence at a specific timescale, beyond what is explained by seizure occurrence alone.

We then investigated if SNS duration also varied over spike rate cycles. For seizures with a given SNS, we computed the non-parametric circular-linear correlation $D$[45] between the SNS duration and the phases of the spike rate cycle at which the seizures occurred, using permutation tests to determine statistical significance (see Suppl. Methods). The measure $D$ ranges from 0 to 1, with zero indicating no association.

Fig. 6A-D shows an example SNS duration association with a spike rate cycle in NeuroVista 11. In this patient, the circadian spike rate cycle was significantly associated with the duration of SNS A after FDR correction for multiple comparisons ($p = 0.0001$). SNS A's duration was markedly higher during the rising phase than during the falling phase of the spike rate cycle (Fig. 6C); the average duration starts increasing shortly before the trough

of the cycle and peaks at approximately $3\pi/2$ in the rising phase before decreasing again (Fig. 6D).

In our cohort, seven of the ten patients had one or more SNSs with significantly associated spike rate cycle phases in terms of duration (Fig. 6D-E). Of these patients, three had circadian associations and five had multidien associations (Fig. 6F). The strength of the correlations between SNS duration and spike rate cycle phases varied from 0.04 to 0.18 (median: 0.09) (Fig. 6G). As with SNS occurrence, we interpret these associations as evidence that SNS duration can be modulated over the timescales of spike rate cycles.

Finally, to conclude our analysis, we investigated if SNS duration and occurrence are modulated by the same temporal factors (Supplementary section S8). We analysed coincidence of SNS duration and occurrence modulation by the same factor, such as the same spike rate cycle. Supplementary section S8 shows that this coincidence rate was relatively low and not above chance-level.

## Discussion

We analysed variability in seizure network states (SNSs) in chronic iEEG recordings, providing novel insight into the patterns and mechanisms of seizure variability. We found that in most patients, SNSs depended on when the seizure occurred in the recording, with some SNSs becoming more or less prevalent and/or increasing or decreasing in duration as the recording progressed. Additionally, several patients had one or more SNSs associated with circadian and/or multidien cycles in interictal spike rate. These associations suggest that seizure features are modulated over multiple timescales, including circadian and multidien timescales that can be revealed by interictal biomarkers.

We first found that seizure evolutions often depended on the amount of time elapsed since the start of the recording. Variability over multiple months to years may reflect non-cyclical changes due to factors such as post-implantation effects[17,31,32], medication changes[46], and slow changes in the epileptic network due to plasticity[47]. Analysing longer recordings in future could also determine if persistent seizure variability reflects longer cycles, such as circannual cycles, in brain dynamics[35,36,48]. Notably, transiently observed SNSs at the beginning of recordings were uncommon in our cohort, suggesting that implantation effects rarely cause atypical seizure network SNSs. Thus, shorter presurgical recordings patients with epilepsy, which typically last for a few days to a few weeks, likely contain a patient's usual SNSs, although SNS duration and relative SNS prevalence may change over time. Our findings add to the existing literature on variability in brain dynamics across chronic iEEG recordings[17,31,32] by revealing that multiple features of seizure evolutions also vary across these longer timescales.

Our work builds on past research that provided evidence for seizure variability over specific timescales. For example, it is well-established that in some patients, clinical seizure features such as focal to bilateral tonic-clonic spread are associated with sleep/wake state or day/night cycles[12–16]. Past analysis of chronic iEEG in canines also discovered shifts in seizure onset patterns as the recording progressed, likely due to postimplantation variability in brain dynamics[17]. Additionally, variability in seizure onset and spread have been linked to preictal and interictal changes in network features[22], band power[49], the location of high frequency oscillations[3], and patterns of cortical excitability[50]. The same interictal features (network dynamics[51,52], band power[53], high frequency activity[31], and cortical excitability[54]) have all been shown to vary over circadian and/or multidien cycles. We now show that seizure evolutions also change over the timescales that influence interictal brain dynamics, suggesting that these fluctuations share common mechanisms. These exact mechanisms are still elusive, and it is also unclear if circadian mechanisms are truly independent or more tied to sleep/wake cycles[36].

Although we found many associations between seizure timing, spike rate cycles, and SNSs, we were unable to explain the full spectrum of SNS variability in our patients. Other approaches could yield more comprehensive and stronger explanations of seizure features. First, our analysis focused on spike rate phase due to its association with seizure occurrence[34–36]. However, we also observed that the amplitude of spike rate cycles often varied over time, potentially reflecting variability in the strength of these cycles. Such changes in cycle strength could potentially impact seizure features. Second, as with seizure occurrence[34,35,37], different cycles likely interact to produce the observed seizure variability. A predictive model incorporating multiple timescales may be more informative than a single spike rate cycle[53]. Third, we limited our analysis to each patient's overall spike rate. Spatial patterns of spike rate also vary over time[31], and other interictal events such as high frequency activity have different temporal profiles than spike rate[31]. Spatiotemporal variability in interictal dynamics may be linked to seizure variability[3] and spatial patterns of different interictal events could also be incorporated in multivariate models of seizure features. Other approaches for analysing SNS could also uncover additional modulations; in particular, although a given SNS's occurrence and duration were often independently modulated, there were likely interactions across SNSs. For example, the duration of an early SNS may depend on whether a seizure progresses to a subsequent possible SNS[55]. Approaches such as canonical correlation analysis could uncover combinations of seizure features that are associated with combinations of interictal features. Furthermore, for our analysis of SNS occurrence we asked if the modulation was significantly different to the overall seizure occurrence modulation, thus making it more difficult to detect any weak SNS modulations that followed the same pattern as the seizure occurrence. Finally, we focused on SNSs, as measured by functional connectivity of the iEEG. Other seizure features may reveal more or additional modulations, which we hope future work will reveal.

Understanding patient-specific seizure variability could provide new clinical strategies for managing seizures in patients with focal epilepsy. First, cycles in seizure features could be added to seizure forecasting algorithms,[56–58] allowing them to forecast not only when a seizure will occur, but also how the seizure will manifest. Additionally, clinicians could modify a patient's antiepileptic medication based on both seizures likelihood[57–59] and seizure severity.[60] Both interictal and seizure variability may also have implications for treatment efficacy; for example, in a mouse model of temporal lobe epilepsy, optogenetic stimulation only impacted seizures that arose from specific brain states[10]. Novel, seizure-specific treatments could therefore be designed to fluctuate over the same timescales as the patient's seizures, thus delivering time-adaptive treatments that account for the patient's seizure variability. Finally, uncovering the time-varying mechanisms that underlie seizure variability and severity could provide new targets for manipulating seizures and lessening their impact on patients.

While the NeuroVista dataset provides an uncommon opportunity to explore seizure variability over longer timescales, it has some limitations. First, due to its small sample size, we were unable to investigate whether there were consistent patterns in seizure modulation across patients. Repeating our analysis in a larger cohort could determine if there are certain characteristic timescales of seizure variability, analogous to chronotypes in seizure occurrence[15,35,48,61]. Likewise, due to our small sample size and heterogeneous findings, we were unable to compare patient clinical and demographic features, such as age, sex, pathology, or seizure onset zone location, to variability patterns. We also lacked clinical information at the seizure level, such as seizure severity or semiology, that could be related to SNSs and variability patterns. However, other studies have related seizure networks to clinical seizure types[9,20–22]. Thus, some SNSs likely relate to seizure clinical characteristics, and future studies could use clinical seizure data or quantitative seizure severity markers[62] to explore this relationship and investigate how seizure symptoms fluctuate over time. Finally, other datasets, such as presurgical iEEG recordings, provide better spatial coverage as well as accompanying neuroimaging data. Such data would provide opportunities to relate seizure networks to the patient's underlying structural connectivity[63].

In summary, we have shown that features of seizure evolutions vary over multiple timescales within individual patients with focal epilepsy. Like interictal dynamics, seizures can change over the months and years of chronic iEEG recordings as well as over faster timescales, such as circadian and multidien cycles. As with cycles in seizure occurrence, cycles in seizure features can be extracted using interictal spike rate as a biomarker. Future work could explore whether fluctuations in other interictal features, such as spatial patterns of spikes and high frequency activity, explain additional seizure features. Ultimately, uncovering the timescales of within-patient seizure variability could lead to new time-adaptive approaches for controlling seizures.

## Acknowledgements

We thank members of the Computational Neurology, Neuroscience & Psychiatry Lab (www.cnnp-lab.com) for discussions on the analysis and manuscript; P.N.T. and Y.W. are both supported by UKRI Future Leaders Fellowships (MR/T04294X/1, MR/V026569/1).

# Figures and captions:

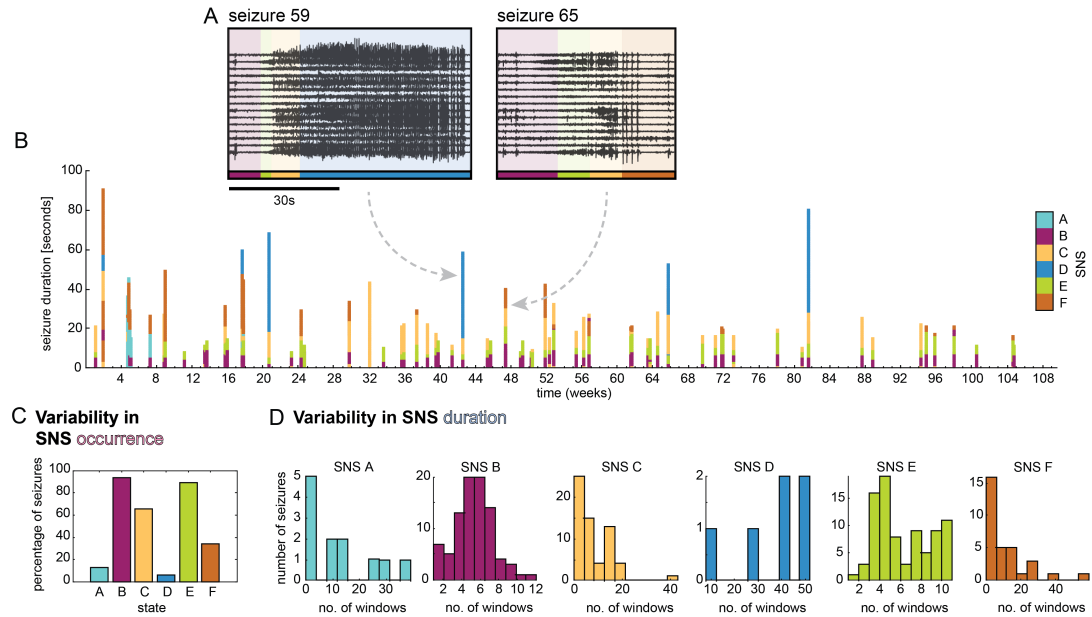

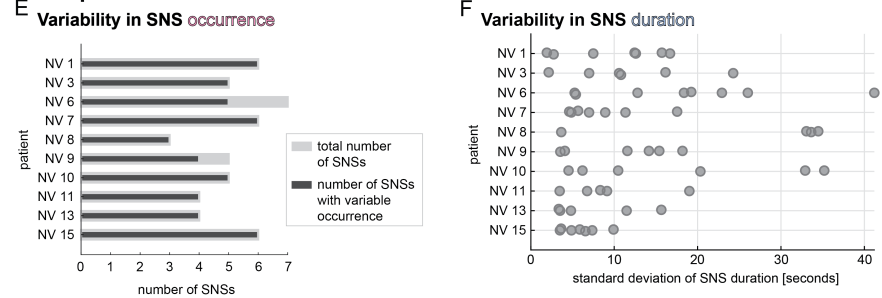

***Variability in seizure network state (SNS) progressions.*** A-D) Example patient, NeuroVista 1: A) Intracranial EEG of two example seizures, with their corresponding overlaid. B) All seizures: each seizure is represented by a vertical bar; its horizontal location indicates the time of seizure occurrence and its height indicates the seizure's duration. The colours of the bar indicate the SNS. C) Percentage of seizures that contained a given SNS in NeuroVisa 1. D) Histograms of SNS duration in NeuroVisa 1. Seizures that do not contain a given SNS are excluded from the corresponding histogram. E-F) All patients: F) The number of SNSs in each patient (light gray bars), with the number of those SNSs that had variable occurrence (i.e., did not occur in all of the patient's seizures) overlaid with dark gray bars. G) Variability in SNS duration for each patient. Each dot corresponds to a single SNS. NV = NeuroVista.

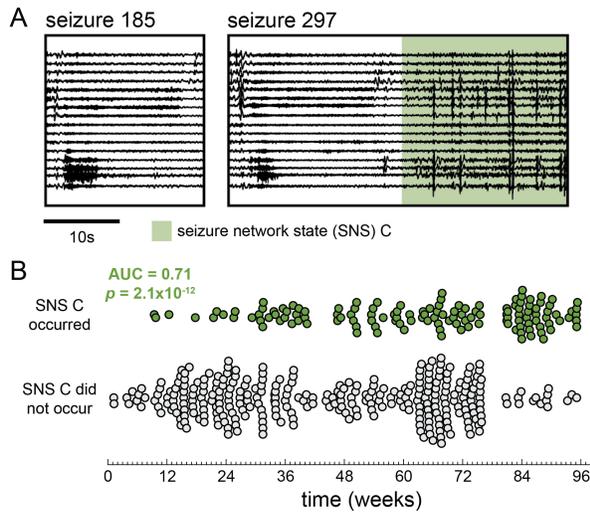
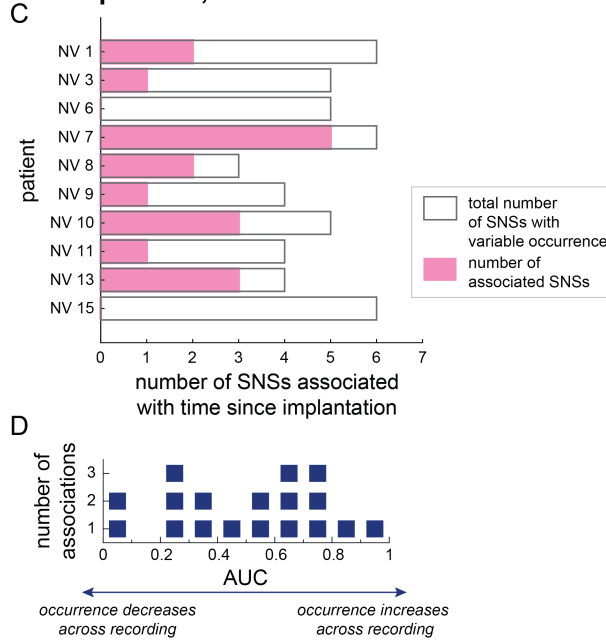

***Relationship between seizure network state (SNS) occurrence and time since implantation.*** *A-B) Example patient, NeuroVista 13: A) Example iEEG of seizures with and without SNS C (green). Both seizures begin similarly, but only seizure 297 progresses further to SNS C. B) Time since implantation of seizures with (green circles) and without (grey circles) SNS C. Points are spread vertically to prevent overlap between seizures with similar times. Time since implantation separated seizures with and without SNS C with an AUC of 0.71. C-D) All patients: C) Number of SNSs with significant associations with time since implantation in each patient (pink bars). Grey outlines provide a reference for the maximum possible number of associated SNSs (i.e., the number of SNSs with variable occurrence, equivalent to the dark grey bars in Fig. 1F). D) Dot plot of the AUCs for each significant SNS in (C). Each blue marker corresponds to one SNS.*

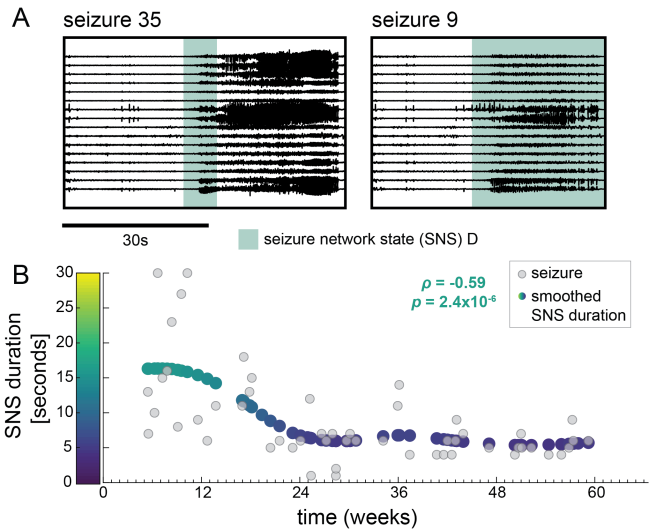
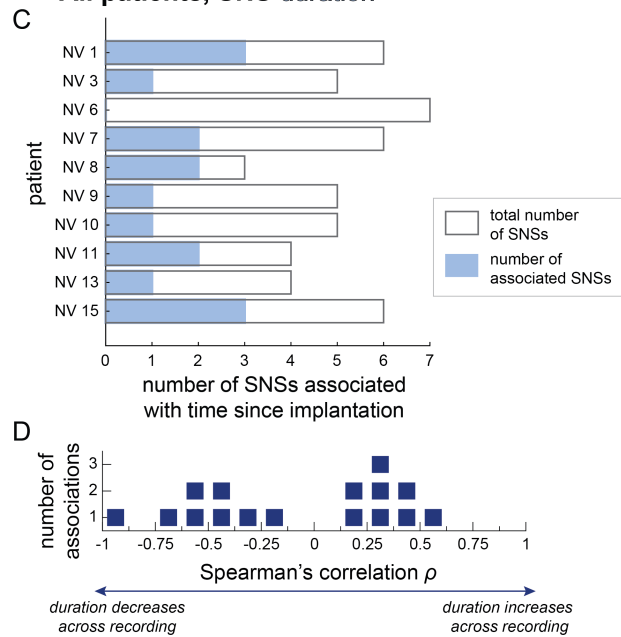

***Relationship between seizure network state (SNS) duration and time since implantation.*** *A-B) Example patient, NeuroVista 15: A) Example iEEG of seizures with SNS D (teal). Seizure 35 briefly visits SNS D before progressing to another SNS, while seizure 9 lingers in SNS D before terminating. B) SNS D duration versus the time since implantation, with a smoothed trend line (Gaussian window of 24 weeks) shown with the coloured points. Each grey point corresponds to a seizure that contained SNS D. C-D) All patients: C) Number of SNSs with significant associations with recording time in each patient (light blue bars). Grey outlines provide a reference for the maximum possible number of associated SNSs (i.e., the total number of SNSs in each patient, equivalent to the light grey bars in Fig. 1E). D) Dot plot of the Spearman's correlation between SNS duration and recording time for all significantly associated SNSs. Each blue marker corresponds to one SNS. For each SNS, we*

*only analysed SNS duration in seizures containing the SNS, thus ensuring that our results were not driven by seizures without the SNS.*

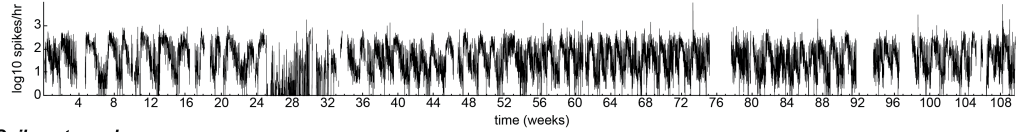
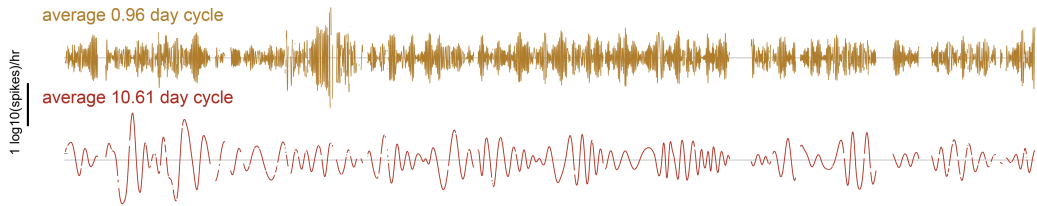
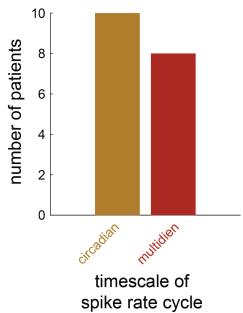
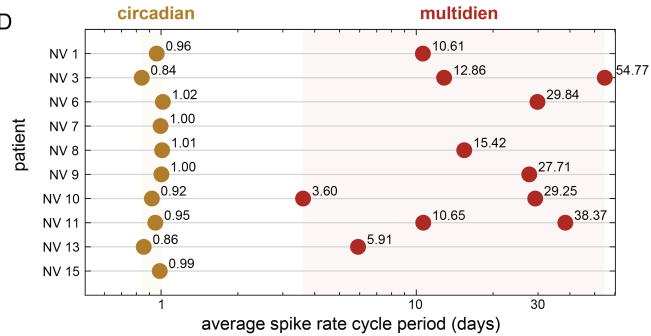

**Patient-specific cycles in interictal spike rate.** *A-B) Example patient Neurovista 1: A) The interictal spike rate (number of spikes per hour versus recording time) during NeuroVista 1's recording. B) The two prominent cycles in NeuroVista 1's spike rate, extracted using empirical mode decomposition (EMD). NeuroVista 1 had a circadian (average period of 0.96 days) and multidien (average period of 10.61 days) cycle. C-D) All patients: C) Number of patients that had at least one spike rate cycle at each timescale. D) Spike rate cycles of each patient, coloured by their timescale.*

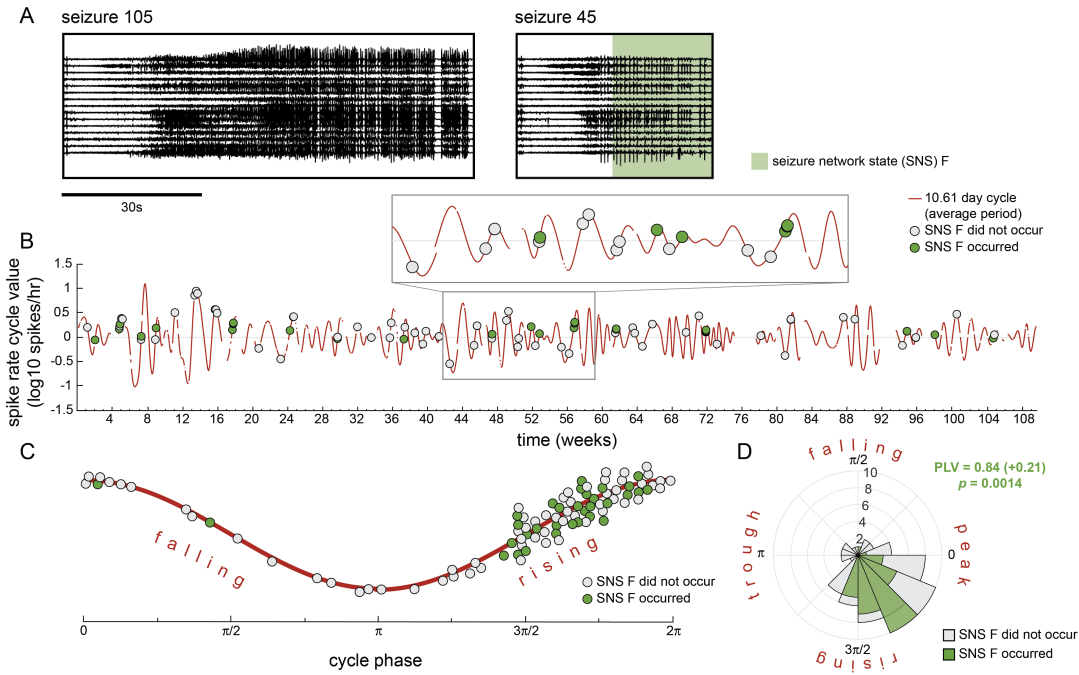

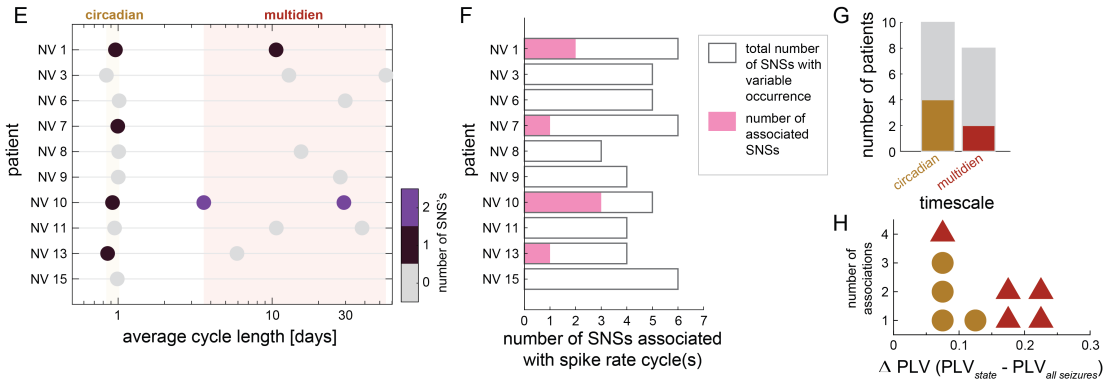

***Associations of seizure network state (SNS) occurrence with spike rate cycles*** *NV = NeuroVista. A-D) Example patient NeuroVista 1, SNS F, 10.61 day cycle: A) Example iEEG of seizures with (seizure 105) and without (seizure 45) SNS F (green). B) The 10.61 day spike rate cycle shown in time, with seizures indicated as circles. Circles are coloured by whether the seizure did (green) or did not (grey) contain SNS F. C) The 10.61 day spike rate cycle shown in phase and seizures indicates as before. D) Polar histogram of seizure phases with (green) and without (grey) SNS F. The statistical significance of SNS F's PLV was determined using a permutation test that randomised which seizures contained SNS F. E-H) All patients: E) Coloured circles indicate the spike rate cycles with significantly associated with one or more SNSs, with statistical significance determined using permutation tests as in (D). F) Total number of SNSs associated with spike rate cycles in each patient. Grey outlines provide a reference for the maximum possible number of associated SNSs. G) Summary of patients with circadian or multidien associations. Grey bars indicate the number of patients with each timescale (equivalent to the coloured bars in Fig. 4C).*

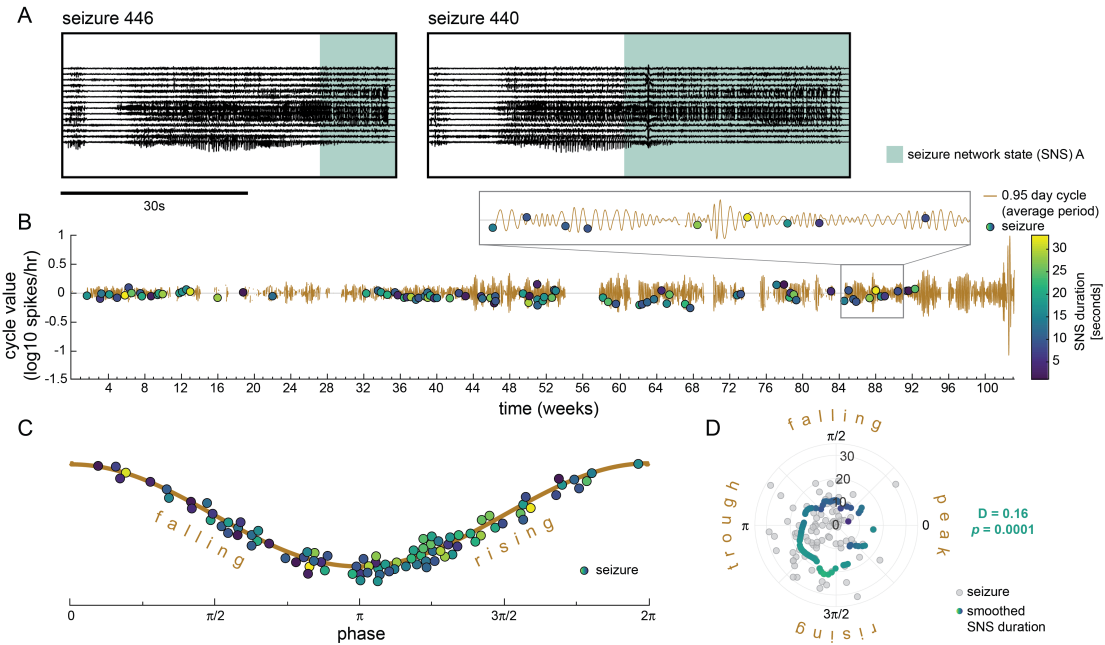

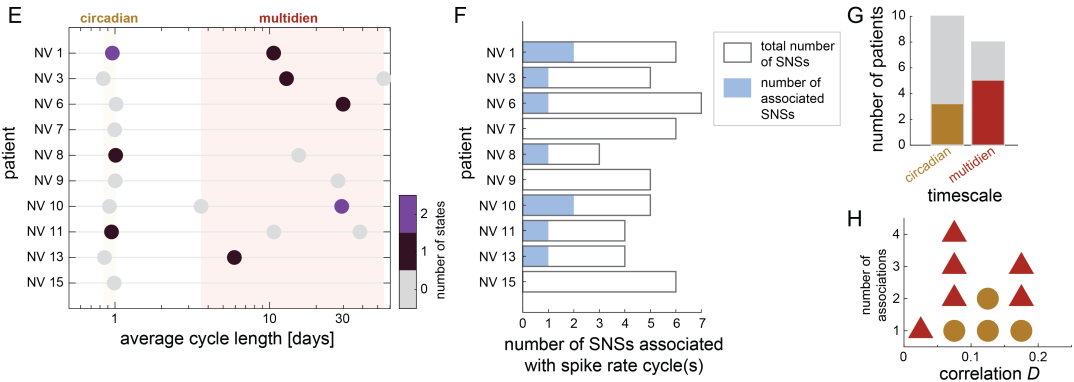

***Associations of seizure network state (SNS) duration with spike rate cycles.*** *NV = NeuroVista. A-D) Example patient NeuroVista 11, SNS A, 0.95 day spike rate cycle: A) Example iEEG of seizures containing SNS A with varying duration. B) The circadian spike rate cycle shown in time, with seizures containing SNS A (circles) indicates as circles coloured by SNS duration. C) Representation of the phases of the spike rate cycle. As in (B), circles indicate seizures with SNS A, and their colour indicates SNS duration. D) Polar scatter plot of SNS duration versus seizure phase. Each grey point indicates a seizure, while coloured points correspond to the smoothed seizure duration (π/4 Gaussian window). The significance of the rank linear-circular correlation D between SNS A duration and seizure phases was determined using a permutation test that randomly shuffled the seizures' SNS A duration. E-H) All patients: E) Coloured circles indicate spike rate cycles with one or more significant associations in SNS duration, with significance determined by a permutation test as in (D). F) Total number of SNSs that were associated with spike rate cycles in terms of duration in each patient. Grey outlines provide a reference for the maximum possible number of associated SNSs. G) The number of patients that had at least one SNS duration associated with a spike*

*rate cycle at each timescale. Grey bars indicate the number of patients with each timescale. H) Dot plot of the strengths of the significant associations between SNS duration and spike rate cycle phases, measured as non-parametric circular-linear correlation D. Marker shape and colour indicates the timescale category of the associated spike rate cycle*

## Supplementary Methods

### Seizure iEEG preprocessing

NeuroVista seizure data was previously notch filtered at 50 Hz during the iEEG acquisition and then bandpass filtered (2nd order, zero-phase Butterworth filter from 1-180 Hz) by[38]. After removing any electrodes with noisy or intermittent signal from the analysis, we re-referenced all iEEG to a common average reference. Time periods with signal dropouts were detected using line length and marked as missing data as in[8] (See Supplementary section S7).

### Computing seizure time-varying functional connectivity

Seizure functional connectivity was defined as band-averaged coherence in six frequency bands: delta 1-4 Hz, theta 4-8 Hz, alpha 8-13 Hz, beta 13-30 Hz, gamma 30-80 Hz, high gamma 80-150 Hz. The time-varying coherence in each frequency band was computed for each seizure from onset to termination using a sliding window (10s window, 9s overlap) as in[9]. For each 10s window, the band-averaged coherence was calculated using Welch's method (2s window, 1s overlap, yielding a total of nine subwindows per 10s window). To tolerate some missing data in each seizure, we allowed functional connectivity in each 10s window to be estimated using a subset of the overlapping 2s Welch subwindows. If a 10s time window had five or more 2s subwindows that contained missing data, the seizure was removed from the analysis.

The upper-triangular elements of each symmetric coherence matrix were re-expressed as vectors of length $(n^2 - n)/2$, where $n$ is the number of iEEG channels, and each vector was normalised to have an $L1$ norm of 1. Seizure time windows therefore had $6 \times (n^2 - n)/2$ features describing the pairwise channel interactions in the six different frequency bands.

### Computing progressions of seizure network states (SNS)

To extract seizure network states (SNSs), patterns of recurring functional connectivity were identified in each patient by applying stability NMF[39,40] to all of a patient's seizure functional connectivity time windows using the same pipeline as in our previous work[9]. This step described each patient's time-varying seizure functional connectivity using (1) a small number of patient-specific NMF basis vectors that captured patterns of functional connectivity, and (2) time-varying coefficients that denoted the contribution of each basis vector to each time window's connectivity.

We observed that most seizure time windows had a single NMF basis vector with a high coefficient. As such, a time window's dominant basis vector (i.e., the basis vector with the highest coefficient) provided a simplified description of the time window's functional connectivity. Therefore, a seizure's time-varying functional connectivity could be simplified as a sequence, or progression, of SNSs, where the SNS of each time window was the dominant NMF basis vector. We used this approach to describe each patient's seizures as progressions of SNSs.

## Preprocessing of interictal spike rate

Detection of interictal epileptiform spikes for this dataset was previously performed and validated[41]. The time-varying spike rate for each patient's recording was summarised as the number of spikes, across all channels, in non-overlapping one hour windows. Due to communication dropouts or failures to regularly store the iEEG data, each one hour segment could have missing segments that affected the spike rate count. To normalise for these dropouts, we normalised each hourly spike rate count by the proportion of captured iEEG data:

$$S_{t,norm} = \frac{S_{t,rec}}{1 - D_t}$$

where $S_{t,rec}$ is the recorded spike rate (spikes/hr) of hour $t$, $S_{t,norm}$ is the spike rate (spikes/hr) of hour $t$ after normalising for dropouts, and $D_t$ is the proportion of dropout time, or missing data, of hour $t$. Recording hours with $D \geq 0.75$ were considered missing data.

After normalising for dropouts, each hourly spike rate $S_{t,norm}$ was log transformed:

$$S_t = log10(S_{t,norm} + 1)$$

yielding the final spike rate, $S_t$, of each hour $t$.

Beginning with the shortest missing segments of spike rate data, we then iteratively imputed missing segments of spike rate data using a method similar to[34]. For each missing segment, we first selected the spike rate data segments directly preceding and following the missing segment that were the same length as the missing segment. If this data was available (i.e., did not contain missing values or exceed the endpoints of the recording), we used the surrounding segments to generate spike rate data for the missing segment. This data was generated by linearly interpolating between the means of the surrounding segments and then adding Gaussian noise with a mean of zero and standard deviation of the surrounding segments. Any resulting interpolated data with a spike rate of less than zero was changed to zero. The length of each interpolated segment was also recorded so that interpolated data was only included in the analysis when its length was much shorter ($\leq 20\%$) than the period of the analysed spike rate cycle (extracted in Methods section 12.6). Any remaining missing time points were temporarily set to the mean spike rate value prior to EMD and then returned to missing values after the decomposition and Hilbert transform (Methods section 12.6).

## Extracting interictal spike rate cycles using empirical mode decomposition (EMD)

Empirical mode decomposition (EMD)[42] was then used to extract intrinsic mode functions (referred to as "spike rate cycles" in the Results) from each spike rate time series. Briefly, EMD decomposes any given signal into a set of signals known as intrinsic mode functions (IMFs) that exactly reconstruct the original signal when summed together along with a with a residual signal. A key property of each IMF is that it must have approximately the same number (up to +/- 1) of extrema as zero-crossings, which ensures that there are no riding waves in the extracted IMFs, as well as a local mean (i.e., the mean of the maximal and minimal envelopes of the IMF) of zero. This property also ensures a well-defined Hilbert transform, allowing us to extract the phase of the signal fluctuations.

We used a variation of EMD known as CEEMDAN[43] that helps ensure that each IMF contains oscillations with a similar timescale (i.e., the mode's period does not dramatically vary over time) by adding noise to the time series prior to the decomposition. The standard deviation of the added noise was scanned from 0.0025 to 0.125 in steps of 0.0025, and the decomposition at each noise level was performed with 100 noise realisations, a maximum of 1000 sifting iterations to extract each mode, and the signal-to-noise ratio increasing for every stage of the decomposition. This initial step yielded 50 versions, one for each noise level, of the EMD decomposition for each patient's spike rate time series.

For each decomposition, we used the Hilbert transform to determine the time-varying frequency, phase, and amplitude of each extracted spike rate IMF[42]. We initially estimated the average period of each IMF using the median frequency of only the original (i.e., non-interpolated) spike rate data, excluding the first and last ten days of the recording due to possible instability in the frequency estimate at the time series boundaries. For each IMF, segments with interpolated spike rate were removed if their duration exceeded 20% of the IMF's period. To define each IMF's timescale, we then recomputed the average period of each IMF as above, now using all of the IMF's non-missing data. The average amplitude of each IMF was computed using the same process.

For each noise level, we then computed the pairwise index of orthogonality, $O$[42], between all pairs of time series from the EMD decomposition (i.e., the IMFs and the residue signal):

$$O_{i,j} = \frac{1}{T} \sum_{t=1}^{T} \frac{C_i(t) C_j(t)}{C_i(t)^2 + C_j(t)^2}$$

where $C_i(t)$ is the $i$th extracted time series, $C_j(t)$ is the $j$th extracted time series, $t$ is the time point in each time series, and $T$ is the total number of time points with spike rate data in both time series. The normalisation by $\frac{1}{T}$ allowed us to compare $O$ across pairs of time series that had different amounts of missing data due to the spike rate interpolation step. $O$ is close to zero when the two time series are locally orthogonal (i.e., do not contain oscillations at similar frequencies during the same time interval). Thus, to minimise overlap in the frequencies of different spike rate cycles, we found the maximum absolute value of the pairwise $O$ for each decomposition and then selected the noise level that minimised this value. This decomposition was used for all downstream analysis.

The median amplitudes versus median periods of the IMFs of the selected decompositions are shown in Supplementary section S3. To focus our analysis on the primary, robust contributors to spike rate cycles, we limited our analysis to IMFs with locally prominent amplitudes that had median periods that were less than a quarter of the duration of the patient's recording (see Supplementary section S3). Across patients, we observed a clear distinction between cycles with median periods of approximately one day (0.83 to 1.03 days) and cycles with longer periods (3.93 to 54.77 days). We labeled these timescales as circadian and multidien cycles, respectively.

## Supplemental comparisons

In addition to the main text statistical analysis, we also compared (1) the overall spike rate, prior to EMD, to SNS occurrence and SNS duration using Wilcoxon rank sum tests and Spearman's correlation, respectively, and (2) overall seizure duration to time since implantation and spike rate cycles using Spearman's correlation and rank circular-linear correlation, respectively. These tests were included in the overall FDR correction for multiple comparisons (see main text Methods, "Statistical analysis") and the results are presented in Supplementary section S5.

**References**

1.	Noachtar S, Peters AS. Semiology of epileptic seizures: A critical review. *Epilepsy and Behavior*. 2009;15:2-9. doi:10.1016/j.yebeh.2009.02.029

2.	Jiménez-Jiménez D, Nekkare R, Flores L, et al. Prognostic value of intracranial seizure onset patterns for surgical outcome of the treatment of epilepsy. *Clinical Neurophysiology*. 2015;126:257-267. doi:10.1016/j.clinph.2014.06.005

3.	Gliske SV, Irwin ZT, Chestek C, et al. Variability in the location of high frequency oscillations during prolonged intracranial EEG recordings. *Nature Communications*. 2018;9:2155. doi:10.1038/s41467-018-04549-2

4.	King-Stephens D, Mirro E, Weber PB, et al. Lateralization of mesial temporal lobe epilepsy with chronic ambulatory electrocorticography. *Epilepsia*. 2015;56(6):959-967. doi:10.1111/epi.13010

5.	Salami P, Peled N, Nadalin JK, et al. Seizure onset location shapes dynamics of initiation. *Clinical Neurophysiology*. 2020;131:1782-1797. doi:10.1016/j.clinph.2020.04.168

6.	Saggio ML, Crisp D, Scott J, et al. A taxonomy of seizure dynamotypes. *Elife*. 2020;9:e55632. doi:https://doi.org/10.7554/eLife.55632

7.	Cook MJ, Karoly PJ, Freestone DR, et al. Human focal seizures are characterized by populations of fixed duration and interval. *Epilepsia*. 2016;57(3):359-368. doi:10.1111/epi.13291

8.	Schroeder GM, Chowdhury FA, Cook MJ, et al. Multiple mechanisms shape the relationship between pathway and duration of focal seizures. *Brain Communications*. 2022;4(4). doi:10.1093/braincomms/fcac173

9.	Schroeder GM, Diehl B, Chowdhury FA, et al. Seizure pathways change on circadian and slower timescales in individual patients with focal epilepsy. *Proceedings of the National Academy of Sciences*. 2020;117(20):11048-11058. doi:10.1073/pnas.1922084117

10.	Ewell LA, Liang L, Armstrong C, Soltész I, Leutgeb S, Leutgeb JK. Brain state is a major factor in preseizure hippocampal network activity and influences success of seizure intervention. *The Journal of Neuroscience*. 2015;35(47):15635-15648. doi:10.1523/JNEUROSCI.5112-14.2015

11.	Ryzi M, Brazdil M, Novak Z, et al. Long-term outcomes in patients after epilepsy surgery failure. *Epilepsy Research*. 2015;110:71-77. http://www.embase.com/search/results?subaction=viewrecord{\&}from=export{\&}id=L601509592{\%}5Cnhttp://dx.doi.org/10.1016/j.eplepsyres.2014.11.011

12.	Sinha S, Brady M, Scott CA, Walker MC. Do seizures in patients with refractory epilepsy vary between wakefulness and sleep? *Journal of Neurology, Neurosurgery and Psychiatry*. 2006;77:1076-1078. doi:10.1136/jnnp.2006.088385

13. Bazil CW, Walczak TS. Effects of sleep and sleep stage on epileptic and nonepileptic seizures. *Epilepsia*. 1997;38(1):56-62. doi:10.1111/j.1528-1157.1997.tb01077.x

14. Bazil CW. Seizure modulation by sleep and sleep state. *Brain Research*. 2018;1703:13-17.

15. Loddenkemper T, Vendrame M, Zarowski M, et al. Circadian patterns of pediatric seizures. *Neurology*. 2011;76:145-153. doi:10.1212/WNL.0b013e318206ca46

16. Janz D. The grand mal épilepsies and the sleeping-waking cycle. *Epilepsia*. 1962;3(1):69-109. doi:https://doi.org/10.1111/j.1528-1157.1962.tb05235.x

17. Ung H, Davis KA, Wulsin D, et al. Temporal behavior of seizures and interictal bursts in prolonged intracranial recordings from epileptic canines. *Epilepsia*. 2016;57(12):1949-1957. doi:10.1111/epi.13591

18. Spencer SS. Neural networks in human epilepsy : evidence of and implications for treatment. *Epilepsia*. 2002;43(3):219-227.

19. Kramer MA, Cash SS. Epilepsy as a disorder of cortical network organization. *Neuroscientist*. 2012;18(4):360-372. doi:10.1177/1073858411422754

20. Burns SP, Santaniello S, Yaffe RB, et al. Network dynamics of the brain and influence of the epileptic seizure onset zone. *Proceedings of the National Academy of Sciences*. 2014;111(49):E5321-E5330. doi:10.1073/pnas.1401752111

21. Schindler K, Leung H, Elger CE, Lehnertz K. Assessing seizure dynamics by analysing the correlation structure of multichannel intracranial EEG. *Brain*. 2007;130:65-77. doi:10.1093/brain/awl304

22. Khambhati AN, Davis KA, Lucas TH, Litt B, Bassett DS. Virtual cortical resection reveals push-pull network control preceding seizure evolution. *Neuron*. 2016;91:1170-1182. doi:10.1016/j.neuron.2016.07.039

23. Schindler K, Elger CE, Lehnertz K. Increasing synchronization may promote seizure termination: evidence from status epilepticus. *Clinical Neurophysiology*. 2007;118:1955-1968. doi:10.1016/j.clinph.2007.06.006

24. Kramer Ma, Eden UT, Kolaczyk ED, Zepeda R, Eskandar EN, Cash SS. Coalescence and fragmentation of cortical networks during focal seizures. *The Journal of Neuroscience*. 2010;30(30):10076-10085. doi:10.1523/JNEUROSCI.6309-09.2010

25. Khambhati AN, Davis KA, Oommen BS, et al. Dynamic network drivers of seizure generation, propagation and termination in human neocortical epilepsy. *PLoS Computational Biology*. 2015;11(12):e1004608. doi:10.1371/journal.pcbi.1004608

26. Khambhati AN, Bassett DS, Oommen BS, et al. Recurring functional interactions predict network architecture of interictal and ictal states in neocortical epilepsy. *eNeuro*. 2017;4(1):e0091-16.2017. doi:10.1523/ENEURO.0091-16.2017

27.     Davis KA, Sturges BK, Vite CH, et al. A novel implanted device to wirelessly record and analyze continuous intracranial canine EEG. *Epilepsy Research*. 2011;96(1):116-122. doi:https://doi.org/10.1016/j.eplepsyres.2011.05.011

28.     Cook MJ, O'Brien TJ, Berkovic SF, et al. Prediction of seizure likelihood with a long-term, implanted seizure advisory system in patients with drug-resistant epilepsy: A first-in-man study. *The Lancet Neurology*. 2013;12:563-571. doi:10.1016/S1474-4422(13)70075-9

29.     Howbert JJ, Patterson EE, Stead SM, et al. Forecasting seizures in dogs with naturally occurring epilepsy. *PLoS ONE*. 2014;9(1):e81920. doi:10.1371/journal.pone.0081920

30.     Jarosiewicz B, Morrell M. The RNS system: Brain-responsive neurostimulation for the treatment of epilepsy. *Expert Review of Medical Devices*. 2021;18(2):129-138. doi:10.1080/17434440.2019.1683445

31.     Chen Z, Grayden DB, Burkitt AN, et al. Spatiotemporal Patterns of High-Frequency Activity (80-170 Hz) in Long-Term Intracranial EEG. *Neurology*. 2021;96(7):e1070-e1081. doi:10.1212/WNL.0000000000011408

32.     Ung H, Baldassano SN, Bink H, et al. Intracranial EEG fluctuates over months after implanting electrodes in human brain. *Journal of Neural Engineering*. 2017;14(5). doi:10.1088/1741-2552/aa7f40

33.     Karoly PJ, Goldenholz DM, Freestone DR, et al. Circadian and circaseptan rhythms in human epilepsy: a retrospective cohort study. *The Lancet Neurology*. 2018;17:977-985. doi:10.1016/S1474-4422(18)30274-6

34.     Baud MO, Kleen JK, Mirro EA, et al. Multi-day rhythms modulate seizure risk in epilepsy. *Nature Communications*. 2018;9(88):1-10. doi:10.1038/s41467-017-02577-y

35.     Leguia MG, Andrzejak RG, Rummel C, et al. Seizure Cycles in Focal Epilepsy. *JAMA Neurology*. 2021;78(4):454-463. doi:10.1001/jamaneurol.2020.5370

36.     Karoly PJ, Rao VR, Gregg NM, et al. Cycles in epilepsy. *Nature Reviews Neurology*. 2021;17(May). doi:10.1038/s41582-021-00464-1

37.     Maturana MI, Meisel C, Dell K, et al. Critical slowing down as a biomarker for seizure susceptibility. *Nature Communications*. 2020;11:2172. doi:10.1038/s41467-020-15908-3

38.     Karoly PJ, Kuhlmann L, Soudry D, Grayden DB, Cook MJ, Freestone DR. Seizure pathways: a model-based investigation. *PLoS Computational Biology*. 2018;14(10):e1006403. doi:10.26188/5b6a999fa2316

39.     Lee DD, Seung HS. Learning the parts of objects by non-negative matrix factorization. *Nature*. 1999;401:788-791. doi:10.1038/44565

40.     Wu S, Joseph A, Hammonds AS, Celniker SE, Yu B, Frise E. Stability-driven nonnegative matrix factorization to interpret spatial gene expression and build local gene

networks. *Proceedings of the National Academy of Sciences*. 2016;113(16):4290-4295. doi:10.1073/pnas.1521171113

41.     Karoly PJ, Freestone DR, Boston R, et al. Interictal spikes and epileptic seizures: Their relationship and underlying rhythmicity. *Brain*. 2016;139:1066-1078. doi:10.1093/brain/aww019

42.     Huang NE, Shen Z, Long SR, et al. The empirical mode decomposition and the Hubert spectrum for nonlinear and non-stationary time series analysis. *Proceedings of the Royal Society A: Mathematical, Physical and Engineering Sciences*. 1998;454:903-995. doi:10.1098/rspa.1998.0193

43.     Colominas MA, Schlotthauer G, Torres ME. Improved complete ensemble EMD: A suitable tool for biomedical signal processing. *Biomedical Signal Processing and Control*. 2014;14:19-29. doi:10.1016/j.bspc.2014.06.009

44.     Torres ME, Colominas MA, Schlotthauer G, Flandrin P. A complete ensemble empirical mode decomposition with adaptive noise. In: *2011 IEEE International Conference on Acoustics, Speech and Signal Processing (ICASSP)*.; 2011:4144-4147. doi:10.1109/ICASSP.2011.5947265

45.     Mardia KV. Linear-circular correlation coefficients and rhythmometry. *Biometrika*. 1976;63(2):403-405.

46.     Napolitano CE, Orriols MA. Changing patterns of propagation in a super-refractory status of the temporal lobe. Over 900 seizures recorded over nearly one year. *Epilepsy and Behavior Case Reports*. 2013;1:126-131. doi:10.1016/j.ebcr.2013.07.001

47.     Hsu D, Chen W, Hsu M, Beggs JM. An open hypothesis: Is epilepsy learned, and can it be unlearned? *Epilepsy and Behavior*. 2008;13:511-522. doi:10.1016/j.yebeh.2008.05.007

48.     Rao VR, G. Leguia M, Tcheng TK, Baud MO. Cues for seizure timing. *Epilepsia*. 2020;62(S1):S15-S31. doi:10.1111/epi.16611

49.     Naftulin JS, Ahmed OJ, Piantoni G, et al. Ictal and preictal power changes outside of the seizure focus correlate with seizure generalization. *Epilepsia*. 2018;59:1398-1409. doi:10.1111/epi.14449

50.     Badawy R, Macdonell R, Jackson G, Berkovic S. The peri-ictal state: Cortical excitability changes within 24 h of a seizure. *Brain*. 2009;132:1013-1021. doi:10.1093/brain/awp017

51.     Mitsis GD, Anastasiadou MN, Christodoulakis M, Papathanasiou ES, Papacostas SS, Hadjipapas A. Functional brain networks of patients with epilepsy exhibit pronounced multiscale periodicities, which correlate with seizure onset. *Human Brain Mapping*. 2020;41:2059-2076. doi:10.1002/hbm.24930

52.     Kuhnert MT, Elger CE, Lehnertz K. Long-term variability of global statistical properties of epileptic brain networks. *Chaos*. 2010;20(043126). doi:10.1063/1.3504998

53. Panagiotopoulou M, Papasavvas CA, Schroeder GM, Thomas RH, Taylor PN, Wang Y. Fluctuations in EEG band power at subject-specific timescales over minutes to days explain changes in seizure evolutions. *Human Brain Mapping*. 2022;n/a(n/a). doi:https://doi.org/10.1002/hbm.25796

54. Meisel C, Schulze-Bonhage A, Freestone D, Cook MJ, Achermann P, Plenz D. Intrinsic excitability measures track antiepileptic drug action and uncover increasing/decreasing excitability over the wake/sleep cycle. *Proceedings of the National Academy of Sciences*. 2015;112(47):14694-14699. doi:10.1073/pnas.1513716112

55. Kaufmann E, Seethaler M, Lauseker M, et al. Who seizes longest? Impact of clinical and demographic factors. *Epilepsia*. 2020;61:1376-1385. doi:10.1111/epi.16577

56. Freestone DR, Karoly PJ, Cook MJ. A forward-looking review of seizure prediction. *Current Opinion in Neurology*. 2017;30:167-173. doi:10.1097/WCO.0000000000000429

57. Stirling RE, Cook MJ, Grayden DB, Karoly PJ. Seizure forecasting and cyclic control of seizures. *Epilepsia*. 2021;62(S1):S2-S14. doi:10.1111/epi.16541

58. Baud MO, Rao VR. Gauging seizure risk. *Neurology*. 2018;91:967-973. doi:10.1212/WNL.0000000000006548

59. Ramgopal S, Thome-Souza S, Loddenkemper T. Chronopharmacology of Anti-Convulsive Therapy. *Current Neurology and Neuroscience Reports*. 2013;13:339. doi:10.1007/s11910-013-0339-2

60. Cramer JA, French J. Quantitative assessment of seizure severity for clinical trials: A review of approaches to seizure components. *Epilepsia*. 2001;42(1):119-129. doi:10.1046/j.1528-1157.2001.19400.x

61. Langdon-Down M, Brain WR. Time of day in relation to convulsions in epilepsy. *Lancet*. 1929;213(5516):1029-1032.

62. Gascoigne SJ, Waldmann L, Schroeder GM, et al. A library of quantitative markers of seizure severity. *Epilepsia*. 2023;64(4):1074-1086. doi:https://doi.org/10.1111/epi.17525

63. Shah P, Ashourvan A, Mikhail F, et al. Characterizing the role of the structural connectome in seizure dynamics. *Brain*. 2019;142(7):1955-1972. doi:10.1093/brain/awz125